\documentclass{llncs}
\usepackage{lncsedoc}
\usepackage[final]{graphicx}
\begin{document}
\title{The completeness, computability, and extensibility of quantum theory.}

\author{ Hans H. Diel}

\institute{Diel Software Beratung und Entwicklung, Seestr.102, 71067 Sindelfingen, Germany,
diel@netic.de}
\maketitle

\begin{abstract}
The long lasting discussion on the completeness of quantum theory (QT) has not yet come to an end. The discussion is impeded by the lack of a clear understanding of what makes up the contents of a theory of physics in general and of QT specifically. After such an understanding has been developed, a more precise definition of global properties, such as the completeness, computability, and extensibility of a theory, is possible and necessary. This paper addresses these subjects for theories of physics and, in particular, for QT. 
The basis for the definition of the completeness of a theory is the proposed definition of a "formal causal model of a physical theory". The formal model can be applied to discussions of general attributes, such as completeness, computability, and extensibility.
\end{abstract}


\section{Introduction}

From the very beginning of the development of quantum theory, doubts have been raised regarding the "completeness" of the theory. Despite extensions of the theory and attempts to prove the completeness or the incompleteness of the theory, discussions on the completeness/incompleteness of QT have not yet come to an end. In the initial discussions of the subject, doubts expressed by some QT physicists were accompanied by ideas of how to remove the alleged incompleteness of QT. These ideas mainly centered around the addition of "hidden variables" or similar concepts. Consequently (and unfortunately), attempted proofs in which the addition of hidden variables was not possible or did not help were interpreted as proofs that QT was complete. Other attempted proofs for the completeness of QT transformed the question into its negative and tried to prove that QT is not extensible.  

Amidst all of these discussions, the notions of "completeness of QT" and "extensibility of QT" were used in specific ways, which were apparently not always the same and not in line with the intuitive meaning of the completeness of a theory in physics. This paper attempts to regulate the discussion on the completeness and extensibility of QT by suggesting a possible definition of the "completeness of a theory of physics" in general and of the "completeness of QT" specifically.

For the discussion of the completeness and computability of a theory, it is necessary to consider mathematics, with which these subjects have been precisely defined for mathematical theories since the beginning of the twentieth century (i.e., approximately 1930). Section 2 very briefly describes how completeness and computability are defined in mathematical theories. 
Although the definitions used in mathematics cannot be applied unchanged to physics, some similarities can be established.
One of the lessons from looking at the subject in the context of mathematics is that it is advisable to define first what a specific theory of physics is before an analysis of general properties, such as the completeness and computability of the theory, is attempted.
\\
The main subject of this paper is described in three steps:
\begin{enumerate}
\item  The formal specification of a model of a theory of physics.
\\
Before a formal (e.g., mathematical) evaluation of the general properties of a theory of physics can be attempted, a formal specification of what constitutes the (essential) statements of that theory should be established. Section 3 describes a proposal for the definition of a "formal causal model" of a theory of physics. Discussions regarding the completeness (and other properties) of a theory of physics are then related to possible formal models of this theory (section 4). 
\item Is QT complete?
\\
Based on the definitions given in sections 3 and 4, the general properties (completeness, extendibility, computability) with respect to QT are discussed. To determine these properties for QT, the properties of a possible formal causal model of QT have to be considered. However, certain unsolved QT problems,which are described in section 5, prevent the construction of a complete formal causal model. Nevertheless, in section 6, what the inability to construct a formal model of QT means for the properties (completeness, computability and extensibility) is discussed. 
\item To what extent will the proposals for the removal of QT deficiencies result in improved QT completeness?
\\
Under the assumption that the QT deficiencies described in section 5 represent the incompleteness of QT, in section 7, it is discussed whether some specific proposed theories aimed at a removal of these deficiencies would, in fact, result in a more complete QT.
\end{enumerate}

\section{The mathematical definition of the completeness/incompleteness of a mathematical theory}

 Within mathematical logic, a formal system, such as an axiom system,  is called complete, if it can derive every formula that is true.  
G\"odel published his two Incompleteness Theorems in 1931 (see \cite{Hermes}). The basis for the definition of mathematical completeness/incompleteness is the availability of an axiomatic system (for example, Peano axioms of arithmetic), which represents the mathematical theory.

Within mathematics, a function is computable if it can be programmed on a universal computer (such as a Turing machine) such that the program comes to a halt (and delivers the correct results) for all valid input parameter combinations.

\section{The formal model of a  theory of physics}

A formal model of a physical theory consists of the definition of the state of the physical system, plus the laws of the physical theory. Both specifications, the system state and the laws, have to be specified in some formal language (e.g., mathematics), thereby referring to elementary basic elements of the formal language.
\\
The most popular formal language for the specification of a physical theory involves mathematical equations from established mathematical theories. As will be shown in the following, mathematical equations alone are not sufficient to completely describe a theory of physics. For the description of complex processes, algorithms may have to be specified. More complex compound system state structures require a specification language in addition to mathematical entities, such as vectors, matrices, and tensors.
It is not an objective of this paper to propose a "formal physical theory specification language". Nevertheless, some type of formal language is required for the description of the subjects within this paper. Because the specifications contained in this paper are often given in the form of examples, a mixture of different notations is used.

\subsubsection{The system state}

specifies the totality of objects that are referenced by the laws of a theory.
The object structures and interrelations (e.g., aggregates and composite objects) should be recognizable. Where appropriate, the positioning within space should be specified.
The notation used may include, as many as reasonable, standard mathematical notations (e.g., matrices, tensors, vectors).

\subsubsection{The laws of physics} 
have to be specified in some formal way. For most of the laws of the theories of physics, this does not present a problem because these laws are typically defined in terms of mathematical equations.
\\
Regarding the listing of the laws of physics, the question arises whether the listed law(s) and the system state are intended to specify a complete physical theory and, if not, which parts (i.e., statements) of a physical theory, as described in textbooks, should be reflected in the formal model of the theory? 
The answer to this question can be determined only by an (agreed upon) understanding of what the major objective of a physical theory should be and what the essential contents of a theory of physics typically are. Section 3.1 begins with an assumption of this subject.

\subsubsection{Example:}

Some physicists consider quantum mechanics to be largely derivable from the Schr\"odinger equation. Thus, for quantum mechanics, the laws of physics should contain the  Schr\"odinger equation:
 
 $ -  \frac{\hbar^{2}}{2m}  \frac{\partial^{2}\psi(x,t)}{\partial x^{2}} + V(x,t)  \psi(x,t) = i \hbar  \frac{\partial \psi(x,t)}{\partial t} $.
\\
\\
The system state has to contain the variables that are referenced in the laws of physics:
\begin{verbatim}
System-state :=  
  Space x,
  Time t,     
  Particle-set,
  Potential V. 
\end{verbatim}
\normalsize
Particle-set := $ \psi_{1}, ...,  \psi_{n} $

\subsection{Formal \emph{causal} model of a physical  theory}

Under the assumption that the major objective of physics is the determination of the relationship between the possible states of a physical system, in particular, the determination of the laws which define the possible state(s) that evolve (or may evolve) from a given state $ s_{0} $, the proposed formal definition of a theory of physics focuses on this causal aspect of a physical theory.  

For the definition of a formal causal model of a physical  theory, the laws of physics are represented by a set of conditional statements that define how an "in" state $ s_{0} $  evolves into an "out" state $ s_{1} $ 

    $      L_{1} : \: IF \: c_{1}(s_{0})\: THEN \: s_{1} = f_{1}(s_{0}); $

    $      L_{2} : \: IF \:c_{2}(s_{0}) \: THEN  \: s_{1} = f_{2}(s_{0}); $

            ...

      $     L_{n} : \: IF \:c_{n}(s_{0}) \: THEN   \:s_{1} = f_{n}(s_{0}); $
\\
The "in" conditions $ c_{i}(s_{0}) $ specify the applicability of the state transition function $ f_{i}(s_{0}) $ in basic formal (e.g., mathematical ) terms, or refer to complex conditions that then have to be refined within the formal definition.

The state transition function $ f_{i}(s_{0}) $ specifies the update of state $ s_{0} $ in basic formal (e.g., mathematical ) terms or refers to complex functions that then have to be refined within the formal definition.
\\
To enable non-deterministic theories ("causal" does not imply deterministic) an elementary function 

RANDOM(valuerange, probabilitydistribution)
\\
may also be used for the specification of a state transition function. Details of the RANDOM function are given in appendix A.
\\ 
The set of laws $ L_{1}, ... ,L_{n} $ has to be complete, consistent and reality conformal.
\begin {itemize}
\item complete
\\
A formal causal model of a physical theory is called complete if all possible "out" states are also valid "in" states, where a valid "in" state is defined as 

$ validstate(s) = c_{1}(s)  \vee c_{2}(s)  \vee  .... \vee c_{n}(s)  $.
\\
A further requirement with respect to the completeness of a formal definition of a physical theory is that the conditions
$ c_{1}(s), c_{2}(s),  .... c_{n}(s)  $ and the functions $ f_{1}(s), f_{2}(s),  .... f_{n}(s)  $ are further defined in terms of elementary conditions and operations.
\\
Note: It will be very difficult to prove the completeness of a formal causal model, except for special cases, i.e., when arbitrary states can be shown to be valid (i.e., validstate(s) = TRUE for arbitrary states).

\item consistent
\\
A formal causal model of a physical theory
is called consistent if for all possible "out" states, no more than one "in" condition is true, i.e., 

for all possible "out" states
 s  $   c_{i}(s)  \wedge c_{k}(s) = FALSE $ if $ i \not= k  $. 
\\
Note: Here, the proof for consistency becomes easier if it can be shown that the condition  
" $   c_{i}(s)  \wedge c_{k}(s) = FALSE $  if $ i \not= k  $"  holds true for arbitrary states (not only for possible "out" states).

\item reality conformal
\\
A formal causal model of a physical theory is said to conform to reality if (1) all causal relationships that have been observed in nature can also be shown to be contained in the formal causal theory, and (2) no causal relationships contained in the formal causal theory have been disproved in nature. 
\\
Note-1: A stronger condition that requires that every causal relationship contained in the formal causal theory be verified in nature would be unrealistic. 
\\ 
Note-2: Another condition that requires that every causal relationship contained in the formal causal theory be verifiable in principle in nature may be reasonable.
\\
Note-3: "reality conformal" is a type of requirement that, in contrast to the other two requirements, refers to conditions external to the formal causal model. The requirement has been included to restrict the scope of formal causal models considered in this paper.
\end{itemize}

\subsubsection{Example}

Many areas of physics can be described by starting with a specific Lagrangian. For a description of the causal relationships, i.e., the evolution of the system state, the equation of motion is the 
major law. The equation of motion can be derived from the Lagrangian by using the Euler-Lagrange equation.
\\
The Lagrangian for classical mechanics is
\\ L = V - T with 
\\
$ V=V(x), T = \frac{1}{2} m \dot{x}^{2} $. 
\\
The Euler-Lagrange equation leads to the equation of motion

 $     m \ddot{x} =  \frac{\delta V}{\delta x} $.
\\
The specification of the laws of classical mechanics can be given by a list ($ L_{1}, ... ,L_{n} $ ) that distinguishes different cases or by a single general law. The single general law is
\\
  $  L_{1} $ :  IF  ( TRUE ) THEN   FOR (all Particles $ P_{i} $ ) \{
\\
  $   \;\;  P_{i} = applyEquationOfMotion(P_{i}); $
\\
\}
\\
Thus, the system state has to contain
\begin{verbatim}
System-state := {
  Space,    
  Particle-set,
  Field V := V(x);
}
\end{verbatim}
Particle-set := $ P_{1}, ...,  P_{n} $;  Particle $ P = \{m, x, \ddot{x}, \dot{x} \} $.
 
\subsection{Formal \emph{continuous} causal model of a physical  theory}

The specification of a  causal model of a physical  theory as described in section 3.1 allows for much freedom concerning the distance between the states $ s_{0} $ and $ s_{1} $. This may result in a somewhat approximate or even incomplete specification of 
the causal relationships of a theory. This may be sufficient for certain types of theories, such as statistical theories. 
There are, however, theories in which more precise laws are available and should be reflected in the formal model of the theory. Such models are called "continuous causal models" of physical theories in this paper.

To define a continuous causal model of a physical theory, a "physics-interpreter" is introduced. 
The physics-interpreter acts upon the state of the physical system. The physics-interpreter determines new states continuously in uniform time steps.
For the formal definition of a continuous causal physical theory, the continuous repeated invocation of the physics-interpreter to realize the progression of the state of the system is assumed. 
\\
\\
$ systemstate := \{ spacetimepoint ... \} \\
\; \;   spacetimepoint :=\{  t,  x_{1}, x_{2}, x_{3}, \psi \} \\
\; \;  \psi :=  \{ stateParameter_{1}, ... , stateParameter_{n} \} $
\\
\\
$ systemEvolution( system\; \; S ) := \{    \\
 S.t = 0; S.x_{1}=0; S.x_{2}=0; S.x_{3}=0;  \\
 S.\psi = initialState; \\
 \Delta t = timestep;     ||\;  must\; be\; positiv  \\
 DO \;  UNTIL ( nonContinueState( S ) )  \{ \\
\; \; \; \;       physicsInterpreter(  S, \Delta t );   \\
\}  $
\\
\\
$ physicsInterpreter(  S,  \Delta t ) := \{  \\
\; \;    tdt  = W.t+ \Delta t ;  \\
\; \;    S = applyLawsOfPhysics( S, \Delta t );  \\
\; \;    discardAllSpacetimePointsWithTimeCoordinate( S.t <  tdt ); 
\footnote{The statement $ discardAllSpacetimePointsWithTimeCoordinate( S.t <  tdt ); $ is not of interest in the context of the present paper.}
 \\
\} $
\\
\\
The refinement of the statement $  S = applyLawsOfPhysics( S, \Delta t ); $ leads to the formalism described in section 3.1.:
\\
 $ S = applyLawsOfPhysics( S, \Delta t )  := \{ $
 
    $      L_{1} : \: IF \: c_{1}(s_{0})\: THEN \: s_{1} = f_{1}(s_{0}); $

    $      L_{2} : \: IF \:c_{2}(s_{0}) \: THEN  \: s_{1} = f_{2}(s_{0}); $

            ...

      $     L_{n} : \: IF \:c_{n}(s_{0}) \: THEN   \:s_{1} = f_{n}(s_{0}); $
\\
		$ \} $

A continuous causal model of a physical theory has to satisfy the same requirements for completeness, consistency and reality conformance as described in section 3.1 for causal physical theories.

\subsubsection{Example}

A possible type  of a continuous causal model is the cellular automaton.

In \cite{DielCALagr}, a cellular automaton QFTCA which supports quantum field theory is proposed.
The system state of a QFTCA  system is specified by:
\begin{verbatim}
System-state := {
  Space,    
  Particle-set,
  Field-set.
}
\end{verbatim}
Particle-set := $ \psi_{1}, ...,  \psi_{n} $;  
\\ 
Field-set := $ \phi_{1}, ...,  \phi_{k} $;
\\
\\
At the highest level of specification, the laws of quantum field theory  can be expressed:
\small{
\begin{verbatim}
QFTCA (initial-state)  :=  {
state = initial-state;
DO FOREVER {
   state = state-update-function(state, timestep);
}
}

state-update-function(state, timestep)  :=  {  
FOR ( all fields field[i] ) {
   field-state( field[i] ) = field-update-function(field[i], timestep);
} 
FOR ( all particles pw[k] ) {
   IF ( interaction-occurred( pw[k], pw2 ) THEN
               perform-interaction( pw[k], pw2 );
   ELSE  pw[k] = pw-update-function(pw[k], timestep);	
}
}
\end{verbatim}
\normalsize
For a complete formal model, the functions  field-update-function(), pw-update-function(), interaction-occurred() and perform-interaction() have to refined. Whereas for field-update-function() and pw-update-function(), this can be relatively easily derived from the standard QFT, the refinement of interaction-occurred() and perform-interaction() presents a problem, as will be described in section 5.4.

\section{Properties of physical theories }

\subsection{Complete}

 \emph{A physical theory is called complete if it is possible to construct a formal (continuous) causal model of this theory.}
\\
In section 3 (where the definition of a formal model of a physical theory is described), it is requested that the formal model be complete, consistent and reality conformal.

Note that this definition of a "complete theory of physics" (namely, that it is possible to construct a causal model) is still relatively weak. It does not imply that a "complete theory" cannot or should not be extended.

\subsection{Computable}

A theory of physics is computable if it is possible to provide a formal continuous causal model of this theory and if all the state transition functions  $\{ f_{1}, f_{2} ...  f_{n} \} $ and all the conditions $\{ c_{1}, c_{2} ...  c_{n} \} $ are computable according to the mathematical definition of computability (see section 2). 

If a given formal model of a physical theory contains mathematically non-computable functions, this does not necessarily imply that the underlying (non-formal) theory is not computable. There may be ways to formulate the model of the physical theory such that the usage of mathematically non-computable functions can be avoided.
 
If it is not possible to construct a complete formal model of the physical theory, this is a strong indication that the underlying (non-formal) theory is indeed, in certain areas, non-computable.

\subsection{Deterministic/Nondeterministic}

A formal model of a physical theory is deterministic if no RANDOM function occurs for the state transition functions 
 $\{ f_{1}, f_{2} ...  f_{n} \} $.
A formal model of a physical theory is nondeterministic if at least one RANDOM function occurs for the state transition functions  $\{ f_{1}, f_{2} ...  f_{n} \} $.
\\
Thus, the property deterministic/non-deterministic is primarily a property of the (formal) model of a given theory. It becomes a direct property of the theory if it can be (formally) proven for arbitrary formal models of the theory that the invocation of the RANDOM() function is unavoidable.As will be described in section 6, the author considers the construction of such a proof extremely difficult.

\subsection{Extensible}

There have been attempts to prove the "completeness of QT" by proving that it is not possible to extend QT in such a way that the predictive power of QT will increase  (see \cite{vNeumann} and more recently \cite{Colbeck1} and  \cite{Colbeck2}). 
Increasing the predictive power of QT is thereby understood as making the theory (more) deterministic.
With these attempted proofs for QT completeness, the negated extensibility of QT focuses on the possible addition of further variables ("hidden variables") to the system state. 
Considering the definition of a formal model of a physical theory as described in section 3, the following observations can be made:
\begin{itemize}
\item Extensions of the system state only (e.g., by adding hidden variables) are useless and cannot change a deterministic theory to a nondeterministic theory. It is also necessary to modify some of the laws $\{ L_{1}, L_{2} ...  L_{n} \} $ such that the added variables are also referenced.  
\item The mandatory modification of the laws $\{ L_{1}, L_{2} ...  L_{n} \} $ has to include a replacement of those laws
that include the RANDOM() function to eliminate some or all of their occurrences.
\item Leaving the system state unchanged and attempting to remove the indeterminism via a modification of those places within the theory where the RANDOM() function occurs is at least as promising as adding extra variables.
\end{itemize}

\section{Is QT complete?}

In the following, doubts regarding the completeness of QT are expressed by indicating three deficiencies of QT that are claimed to prevent (at present) the construction of a formal causal model of QT.

\subsection{The measurement problem}

QT consists mainly of the principles, rules, and equations that describe how the probabilities (in the form of probability amplitudes) dynamically evolve in various situations to enable the prediction of the probability of different measurement results. Thus, the ultimate transition to facts due to a measurement is an essential element of QT, although the theory does not say much about when (under which circumstances) and how this transition occurs. Quantum physicists do not seem to consider this lack of explanation as a deficiency of QT, except for possibly agreeing that the measurement problem exists.

The measurement problem of QT can be expressed by a set of questions related to the overall question of what exactly happens during a measurement. The set of questions varies depending on selected basic assumptions to start with. A concise description of the measurement problem is given in \cite{Maudlin} in the form of a trilemma. In \cite{Maudlin}, Maudlin shows that the following three claims are mutually inconsistent:
\begin{enumerate}
\item The wave function of a system  is complete.
\item The wave function always evolves in accord with a linear dynamical equation (e.g., Schr\"odinger equation).
\item Measurements always have a definite outcome.
\end{enumerate}
Maudlin shows variations of these contradicting claims that are contradicting as well. 
\\
The author claims that as long as QT has no generally agreed-upon solution to the measurement problem and this solution is an ordinary part of standard QT, QT should not be considered to be complete. In terms of the above described concept for the completeness/incompleteness of a theory of physics: 
\\
The solution to the measurement problem should be mappable to a formal causal model of QT and thereby the specification of a measurement process should satisfy two conditions:
\begin{enumerate}
\item It consists of ordinary QT/QFT functions only 
\item It does not refer to any parameters, functions or conditions which are outside the scope of QT/QFT (such as, for example, the existence of an "observer", or a condition "capability of determining").
\end{enumerate}

\subsection{"The interference collapse rule"}

When the author began the development of his computer model of quantum theory (QT)  (see \cite{QTModel}), with the goal of supporting the simulation of the major QT (gedanken-) experiments, the double-slit experiment was one of the first samples whose modeling was attempted. Soon it was realized that the double-slit experiment could not be supported by a computer model, at least if the canonical interpretation of the double-slit experiment presented in QT textbooks was used. It turned out (see \cite{Diel1}) that the source of the problem was neither the indeterminism of QT (it is easy to simulate indeterminism), nor the often-mentioned strangeness of QT. The problem was that the typical explanation of the double-slit experiment and the rule for when the interference occurs and when it "collapses" is given in terms of objects and conditions which can neither be transformed to proper mathematics, nor can it be mapped to a facts-driven decision algorithm.

In QT textbooks, the double-slit experiment is often used to explain one of the key principles of QT which in this paper is called the "interference rule". Here, we refer to the formulation of R. Feynman.  The interference rule, in the context of explaining the double-slit experiment, is explained by Feynman in 
\cite {Feynman} as follows:
\\ 
\\
"When an event can occur in several alternative ways, the probability amplitude for the event is the sum of the probability
amplitudes for each way considered separately. There is interference:
\\
$
 \Phi = \Phi_{1}  +  \Phi_{2}
\\ P = |  \Phi_{1}  +  \Phi_{2}  | ^{2}
$
\\
If an experiment is performed which is capable of determining whether one or another alternative is taken, the probability of the event 
is the sum of the probabilities for each alternative. The interference is lost.
\\ $ P = P_{1} + P_{2} $ "
\\
\\
This interference (collapse) rule, or slight variations of it, is contained in all textbooks on basic QT, either as an explanation of the double-slit experiment, or vice versa with the double-slit experiment used to illustrate the rule. Without the photon source near the slits, interference occurs. The inclusion of the photon source aborts the interference.

Currently, when one attempts to map the interference rule to a computer program that has the objective of simulating the double-slit experiment it turns out that this simulation is not possible. The problem is that a condition such as that of an experiment capable of determining whether one or another alternative is taken can neither reasonably be mapped to any mathematical constructs, nor to a physical facts-driven decision algorithm because it does not refer to elements of a physical state, but to some "capability of determining". With a closer examination of  the situation,  it becomes clear that this is not a problem of writing a specific computer program, rather,  the lack of comprehension is more general in nature.
 
In terms of the above described concept for the completeness/incompleteness of QT, the improper interference collapse rule prevents the construction of a complete formal causal model of QT and thus disturbs the completeness of QT.

\subsection{Entanglement}

QT physicists consider Bell's famous
inequality (see \cite{Bell}) and its violation in experiments as a strong indication that "local realistic models" of entanglement are not possible. A model is understood to be a local model, if changes of the state of the system depend on the local state only. Different variants can be found in literature for the definition of "realistic model". In \cite{Laudisa2}  F.  Laudisa describes "realism" as 
"a condition which is often formulated, even recently, as the idea that physical systems are endowed with certain pre-existing properties, namely properties possessed by the systems prior and independently of any measurement interaction and that determine or may contribute to determine the measurement outcomes “.
\\
In section 5.1, a solution to the QT measurement problem is requested to be mappable to a formal causal model which
satisfies the two conditions: 
\begin{enumerate}
\item It consists of ordinary QT/QFT functions,  
\item It does not refer to any parameters, functions or conditions which are outside the scope of QT/QFT.
\end{enumerate}
The requirement of a formal causal model which satisfies these additional conditions implies realism (it is an even stronger requirement).
\\
Whether the present lack of a local realistic causal model of QT entanglement should be viewed as an incompleteness of QT is not clear to the author.  Non-local  realistic causal models of QT entanglement are feasible, though at present not  part of standard QT. 

\subsection{The lack of a functional description}

When R. Feynman explains quantum electrodynamics in  \cite{Feynman1} he writes "I have pointed out these things because the more you see how strangely Nature behaves, the harder it is to make a model that explains how even 
the simplest phenomena actually work. So theoretical physics has given up on that."

Explaining "how phenomena actually work" is referred to in this paper as "a functional description". 
A functional description describes the evolution of a system in terms of state transitions. 
As such, a functional description of a theory of physics may represent a (continuous) causal model of the physical theory described in section 3.2; at least, it should provide a good basis for the construction of a causal model. For most areas of physics, the functional description (or a (continuous) causal model) can easily be derived from the equations of motion of the theory. 
For QT, however, there are three areas in which a functional description is missing because it cannot be derived solely from the equations of motion (e.g., Schr\"odinger equation). The major areas impeding the provision of a functional description of QT are:
\begin{enumerate}
\item The measurement process
\\
In section 5.1, the measurement problem is described as the major deficiency of QT. The author claims that a suitable solution to the measurement problem can be provided by a functional description of the measurement process.
\item The collapse of the interference
\\
In section 5.2, the existing "interference collapse rule" is described as a deficiency of QT. The author claims that a functional description of the interference collapse would remove this deficiency. In \cite{Dieldslit}, an improved interference collapse rule,which equates the interference collapse to the collapse of the wave function, is proposed.

\item The QFT interaction
\\
Quantum field theory (QFT) provides an extensive framework in terms of scattering matrix, Feynman diagrams and Feynman rules for the treatment of interactions (e.g., scatterings) between particles. Nevertheless, the above mentioned statement by Feynman (\cite{Feynman1} "... how even the simplest phenomena actually work ... ") refers to QFT. This is because all these powerful QFT tools do not represent an equation of motion that enables the derivation of a continuous sequence of state transformations.The QFT computations are based on the (idealized) assumption of a time span from 
 $- \infty $ to $+ \infty $.
This is not perceived as a deficiency as long as the details of a QFT interaction are not of interest.
If, however, the QFT interaction is considered to be an important part of each measurement process, more details on the QFT interaction process in the form of a functional description of the QFT interaction may provide a satisfactory explanation of the QT measurement. In \cite{Dielfi}, a functional description of QFT interactions is proposed.

\end{enumerate}

\section{Properties of QT}

The basis for a discussion of QT properties, such as completeness, computability, etc., can only be the theory as it is, at present, generally understood. Discussions on possible extensions of QT to remove possible deficiencies may also be of interest but should be separated from those concerning the present theory.
\\
In section 3 of this paper, the definition of a formal causal model of a physical theory has been proposed with the goal of using this formal definition as the basis for discussions on global properties, such as completeness.This does not mean that for a discussion of, for example, the completeness of QT, a formal causal model of QT has to exist.
If a formal model of QT does not exist and there are convincing arguments for why it cannot be provided, this would be a strong indication (at least to the author) that QT is incomplete. Similar reasoning applies to further global properties, such as computability, determinism, etc.

\subsection{Is QT complete?}

The QT deficiencies described in section 5 prevent the specification of a (complete) formal  continuous causal model of QT (see section 3.2). According to the definition of completeness given in section 4.1, therefore,  \emph{QT is not complete}.

\subsection{Is QT computable?}

Computability is tightly related to the ability to develop an interpreter, as described in section 3.2, for the formal continuous causal model of a theory. Therefore, the inability to provide a (complete) formal continuous causal model of QT also destroys the computability of QT in specific areas.
\\
A concrete example that stresses this conclusion is the fact that some of the QT deficiencies described in section 5 were detected by the author when he attempted to develop a computer model of QT (see \cite{Diel1}).

\subsection{Is QT deterministic?}

At present, QT must be called non-deterministic. With the present understanding of QT, no (complete) formal continuous causal model of QT that avoids the invocation of the RANDOM() function is imaginable.

\subsection{Is QT extensible?}

The author claims that it is very difficult or impossible to prove that QT is not extensible. This holds true for general extensibility, as well as for extensions that would make QT deterministic.

\section{Three attempts to make QT more complete}

Many proposals have been published that attempt to make QT more complete. Many of these proposals have been described under the label "interpretation of QT". Below, three of these proposals are roughly described. The goal of this description is not a detailed introduction to these theories, nor is an evaluation of the theories intended. The primary goal of the description is to show to what extent these theories would enable a more complete construction of a formal causal model of QT.

\subsection{The many-worlds theory}

As a possible solution to the measurement problem, the many-worlds theory (see \cite{Everett}) proposes an alternative to the "collapse" of a wave function by assuming that a wave function will proceed unchanged, but the apparently non-continuing paths will continue within many new worlds. 

As a simplified description, the many-worlds theory could be expressed by a formal causal model of QT in which all occurrences of the RANDOM( ) function (expressing the indeterminism) are replaced by a function

   MULTIPLY\&EXECUTEMANYWORLDS(   ) that expresses the copying and branching into many worlds.
\\ 
This may be viewed as making QT more complete. However, unless further specifications are provided, it does not provide a complete causal model of the process within which at present the RANDOM() functions are embedded.

\subsection{The GRW theory}

The GRW theory (see \cite{GRW}) attempts to offer an explanation of the collapse of the wave function by modifying the (linear) Schr\"odinger equation such that it becomes non-linear and results in reductions and collapses.
If QT physicists could agree on a precise version of the GRW theory, this would enable causal models for the problem areas described in sections 5.1 and 5.2 and thus make QT more complete.

\subsection{The functional model of interactions in quantum theory}

The functional model of interactions in QT described in \cite{Dielfi} has been developed with the goal of offering a solution to the three QT deficiencies described in section 5. The model is based on the assumption that the measurement process and the collapse of the interference (as occurs in the double slit experiment) can be traced back to "normal" QFT interactions and can be explained by a functional model of QT interactions.
\\
The model may be summarized as follows:
\begin{itemize}
\item A measurement is the process through which a measurement apparatus obtains information about the measured object.
\item The measurement process always implies interactions between the measured quantum object and the measurement apparatus.
\\
Measurements of QT observables can be performed using a variety of measurement devices, apparatuses and processes. Common to all such measurement processes is that they have to include at least one interaction in which the measured object exchanges information with some other entity belonging to the measurement apparatus. Such information exchanging interactions are called measurement interactions in the following.
\item Measurement interactions are "normal" interactions that leave behind footprints of the measured object on the measurement apparatus. This may be viewed as a mapping of part of the attributes of the measured object to the object(s) belonging to the measurement device.
In general, the description of a measurement interaction requires QFT, i.e., a scattering matrix, Feynman diagram, etc. 
\item Typically, the exchange of information (i.e., leaving footprints) is accompanied by a reduction and collapse of the wave function of the measured object. Only in this way can the non-definite values associated with the wave function be "reduced" to definite values requested for the measurement results.
\item The functional models system state is described in terms of "particle/wave collections" (pw collection). A pw collection may be thought of as having a two-dimensional structure. In one dimension, the pw collection consists of 1 to n particles (the pw collection contains
$ n > 1 $ particles if these particles are entangled). In the other dimension, the pw collection consists of a number of discrete paths. 
\item each path of a pw collection specifies  definite attribute values for the (1 to n) particles and a 
(single) common probability amplitude.
\item Interactions (such as scatterings and measurement interactions) occur at definite positions. Only the (single) path that covers the interaction position affects the outcome of the interaction and the outcome of a possible measurement. This may be considered to be a reduction of the wave function of the interacting object.
\item All of the paths that do not affect the interaction (because they do not cover the interaction position) will no longer affect the further evolution of the pw collection. This may be viewed as discarding the unused paths or as a "collapse of the pw collection".
\item In general, an interaction (such as scatterings and measurement interactions) results in a set of alternative "out" paths in the form of a (single) "out" pw collection. For QFT, the "out" pw collection may consist of different particles than those contained in the "in" pw collection. 
\item Interactions support only a non-bijective mapping of the "in" state to the "out" state and, thus, only a limited exchange of information. This limited exchange of information is the cause of some of the limitations and peculiarities of QT measurements.
\end{itemize}
Further details on this subject can be found in \cite{Dielmeas}, \cite{Dielfi} and \cite{diel4}.

The functional model of QT contributes much to the provision of a formal causal model of QT because it has been developed with the goal of enabling a causal model. This does not mean that the functional model provides a complete formal causal model of QT.

\section{Discussions}

\subsection{Is it reasonable to focus on causal  laws of physical theories}

The condition for the completeness of a formal model of a theory of physics is relatively weak, insofar as it is not requested that a formal model expresses all  important or major laws of the underlying theory of physics. Completeness of a formal model of a physical theory requires only coverage of the causal laws. This is considered to be reasonable for the following reasons: 
\begin{itemize}
\item It would be rather difficult and probably controversial to specify what the " important" or "major" laws of a theory are.
\item Non-causal laws are often interrelated with causal laws. Typically, the non-causal laws are either (1) derived from the causal laws or (2) vice versa, i.e., the causal laws are derived from the non-causal laws. In both cases, their non-availability in a formal model of a theory of physics should not be understood to be the incompleteness of the model.
\item 	In cases where the causal laws are not interrelated with the non-causal laws, they are often conjectures whose lack should not be considered to be the incompleteness of the formal specification. 
\end{itemize}

\subsection{Can a non-deterministic theory be complete?}

According to the definition of the completeness of a theory (section 4.1) and of a formal causal model of a physical theory, a non-deterministic theory can be complete because inclusion of the RANDOM function within the laws of the theory is valid.
\subsection{Can a complete theory be extended?}

A theory of physics that is complete according to the definitions given in sections 3 and 4 is always extensible. In general, the completeness of a theory does not exclude extensibility.
In several papers, attempts to prove the completeness of QT were undertaken by proving that QT is not extensible. However, in such attempted proofs for the non-extensibility of QT, extensibility was understood based on a very special meaning, namely, extensibility toward a deterministic theory. 
Even if these proofs were successful, according to the definitions given in sections 3 and 4, a proof for the non-extensibility of a theory of physics would not imply the completeness of the theory.




\subsection{Are proofs for QT completeness/incompleteness feasible and reliable?}

Many papers that (a) attempt to prove the completeness of QT, (b) refute the alleged proofs for completeness, or (c) indicate specific QT deficiencies or incompletenesses have been published. In most cases, the proofs took the form of "no-go"-theorems, i.e., the impossibility of something (e.g., extensions of QT) was proven because no-go-theorems are easier to prove than positive predicates. In \cite{Laudisa}, F. Laudisa criticized these attempted no-go-theories by describing them as "... theoretical enterprises ...aiming at establishing negative results for quantum mechanics in absence of a deeper understanding of the overall ontological content and structure of the theory".The author agrees with this criticism and extends it to apply not only to no-go theorems but also to proofs (or analyses) of general properties, such as the completeness, extensibility and computability of theories of physics.

The main subject of this paper is the provision of a formal basis for an "understanding of the overall ontological content and structure of the theory". Thus, proofs for general properties and no-go theorems should become easier, or at least more reliable, if the goal of the paper can be fulfilled. Upon closer examination, it becomes clear that such proofs will become more reliable but not necessarily easier. As shown in mathematics, proofs for the completeness of a theory will be rather difficult if the base theory (represented in terms of the set of axioms) requires more than a few axioms for its formal specification. As G\"odel (see \cite{Hermes}) showed, they are, in some cases, even impossible to achieve. Conversely, the availability of a formal specification of a theory of physics eases proofs (or the analysis) of the incompleteness of QT drastically. It is sufficient to indicate QT deficiencies that prevent the construction of a(complete) formal causal model of QT. Even the non-availability of a formal model of a theory of physics can represent a proof for incompleteness, if it can be shown that the inability to construct a (complete) formal model is due to a certain deficiency of the theory.

\section{Conclusions}

A proper (i.e., formal) determination of the completeness of a theory of physics, such as QT, can only be given if it is possible to provide a complete (formal) model of the subject theory. The provision of a complete model of a theory requires, first of all,
a definition of what the essential content of the theory should encompass. The definition proposed in this paper focuses on the causal laws of a theory of physics, i.e., the ability to predict future system states based on a given present system state. This led to the definition of a formal causal model of a theory of physics (section 3). 

Like all definitions, the definition of the completeness of a theory of physics is not a matter of right or wrong, but rather a matter of applicability and usefulness. The definitions given in this paper may not be suitable for all purposes for all theories of physics.
However, if the general applicability of the proposed definitions is questioned, this strengthens the requirement for a precise definition to be used with discussions on the completeness of QT.

If  it is not possible to provide (at least in principle) a complete formal (causal) model, this can be considered to be a strong indication that the theory is incomplete.
If it is possible to provide a complete formal model, a formal proof for completeness may still be very difficult to achieve. Disproofs, i.e., proofs for incompleteness, are easier because it is sufficient to show a single example that violates the completeness. 
As with the subject completeness, with such further global properties, positive proofs (e.g., for computability) based on the formal model may still be very difficult to achieve. Negative proofs (e.g., for non-computability), however, are easier to achieve and more reliable. 

The incompleteness of QT is argued for because the QT deficiencies described in section 5 prevent the construction of a complete formal causal model of QT. As written in \cite{Ghirardi}: "The issues of the completeness of quantum mechanics, and of the interpretation of the state vector, are by no means resolved".
Proposals for the removal of the QT deficiencies (see section 8), such as the various proposed interpretations of QT, will not automatically achieve the completeness of QT, however, they may at least result in a more complete QT.

%

\appendix

\section{The RANDOM function}

The RANDOM function can be used in the specification of a formal causal model when more detailed specifications
cannot be provided, but rules about the probability distribution can be given. As such, the RANDOM function invocation occurs  within the formal causal model of every non-deterministic theory. It may also occur in the description of 
deterministic theories when detailed (deterministic) specification are not of interest (i.e., within statistical theories).
\\
\\
RANDOM (valuerange, probabilitydistribution)

\begin{itemize}
\item valuerange: the range of possible values that may be returned by the RANDOM function
\item probabilitydistribution: the probability distribution to be used for the random value generation.
\\
Examples: probabilitydistribution  = FLAT, or probabilitydistribution  = GAUSS, or probabilitydistribution  = $ \psi $.
\end{itemize} 

\end{document}